\documentclass{article}
\usepackage{mathrsfs}
\usepackage{amssymb}

\begin{document}

\title{The momentum operators corresponding to a localized massless particle}
\author{
  A.R. Assar and V. Putz$^1$,\\
  {\small$^1$ Institut f\"ur Photonik,
  Technische Universit\"at Wien   }     \\
  {\small 
  Gusshausstr. 25-29,
  A-1040 Vienna, Austria   }            \\
  {\small putz@hep.itp.tuwien.ac.at}}
\maketitle

\begin{abstract}
In this article we propose, using a purely group theoretical argument, that if
a massless particle is localized, then there are only two momentum operator
s corresponding to the localized state. We explicitly determine these self-adjoint operators.
\end{abstract}
\vspace{1cm}
PACS-1996: 02.20.Fh, 03.65.Fd
 
\vspace{1cm}
There have been suggestions \cite{Bahut} that a massless particle can be
localized in the hyperplane perpendicular to the light-like momentum 4-vector of 
that particle. In what follows, we shall use a group theoretical method based
on Mackey's induction-reduction theorem \cite{Mackey}, \cite{Assar}, to obtain
the momentum operators corresponding to such a localized state.

By the relativistic mass-momentum relation,
\begin{eqnarray*}
p^2= p_\mu p^\mu = m^2,
\end{eqnarray*}
if $m=0$, then $p^2=0$, and one may without loss of generality take the light-like vector $p$ to be 
\begin{eqnarray}\label{eq1}
p= \left(\begin{array}{c} 1\\0\\0\\1\end{array}\right).
\end{eqnarray}
There is also a representation of any 4-vector by $2\times2$ matrices as follows
\begin{eqnarray}
P= \left(\begin{array}{cc} p_0+p_3 & p_1+ip_2\\p_1-ip_2& p_0-p_3\end{array}\right).
\end{eqnarray}
Therefore, the matrix $P$ corresponding to the 4-vector (\ref{eq1}) is of the form
\begin{eqnarray*}
P= \left(\begin{array}{cc} 2 & 0\\0& 0\end{array}\right).
\end{eqnarray*}
The little group (or isotropy group) of this 4-vector is the set of all matrices $A$ such that 
\begin{eqnarray}\label{eq3}
A \in SL(2,\mathbb{C}): APA^+ = P.
\end{eqnarray}
Taking 
\begin{eqnarray*}
A= \left(\begin{array}{cc} a & b\\c& d\end{array}\right),
\end{eqnarray*}
it follows from (\ref{eq3}) that 
\begin{eqnarray*}
\left(\begin{array}{cc} a & b\\c& d\end{array}\right)
\left(\begin{array}{cc} 2 & 0\\0& 0\end{array}\right)
\left(\begin{array}{cc} \bar a & \bar c\\\bar b& \bar d\end{array}\right)
= \left(\begin{array}{cc} 2 & 0\\0& 0\end{array}\right)
\end{eqnarray*}
or
\begin{eqnarray*}
\left(\begin{array}{cc} 2|a|^2 & 2a\bar c\\ 2c\bar a & 2|c|^2\end{array}\right)
=\left(\begin{array}{cc} 2 & 0\\0& 0\end{array}\right)
\Rightarrow |a|^2 = 1,\quad c=0.
\end{eqnarray*}
Thus, the isotropy group of $P$ is the set of all matrices in $SL(2,\mathbb{C})$ of the form
\begin{eqnarray*}
IS(P) = \left\{ 
\left(\begin{array}{cc} e^{i\theta} & b\\0& e^{-i\theta}\end{array}
\right)\bigg\arrowvert b\in \mathbb{C}
\right\}.
\end{eqnarray*}
It is clear that this is a subgroup of $SL(2,\mathbb{C})$ which 
is isomorphic to the Euclidean group 
\begin{eqnarray}
E_2 = T_2 \rtimes SO(2).
\end{eqnarray}
The $T_2$-subgroup is clearly the set of matrices 
of the form 
\begin{eqnarray*}
T_2 = \left\{ 
\left(\begin{array}{cc} 1 & b\\0& 1\end{array}\right)\bigg\arrowvert b\in \mathbb{C}
\right\}.
\end{eqnarray*}
The Euclidean $E_2$ is the group of Euclidean motion in a hyperplane
orthogonal to the 4-vector (\ref{eq1}), and the $T_2$ subgroup corresponds to
translations in this hyperplane.

By Wigner's argument particles are classified according to their mass and spin 
by the unitary irreducible representations of the Poincar\'e group, and that
any such representation can be induced from 1-dimensional representations (or
characters) of the little group corresponding to the momentum 4-vector of 
the particle in question. The Poincar\'e group is a semi-direct product
\begin{eqnarray}
{\mathcal P} = T_4\rtimes SL(2,\mathbb{C}),
\end{eqnarray}
where $SL(2,\mathbb{C})$ is the double covering of the restricted Lorentz
group $L_0$. It is this $SL(2,\mathbb{C})$-part, {\it i.e.}, the Lorentz group 
part of the Poincar\'e group which concerns us in what follows. Consider the following two subgroups of 
$SL(2,\mathbb{C})$,
\begin{eqnarray}
&&H_1 = \left\{ 
\left(\begin{array}{cc} \lambda & \mu \\0&
 \lambda^{-1}\end{array}\right)\bigg\arrowvert 
\lambda, \mu \in \mathbb{C}
\right\},\nonumber\\
&&H_2 =\left\{ 
\left(\begin{array}{cc} 1 & \beta \\0& 1\end{array}\right)\bigg\arrowvert \beta\in \mathbb{C}
\right\}.
\end{eqnarray}
Notice that $IS(P)$ is a subgroup of $H_1$. Also $H_1$ is the inducing
subgroup for all principal series unitary IRR of $SL(2,\mathbb{C})$, \cite{Naimark}
.
Our aim is to find the reduction of a unitary representation of
$SL(2,\mathbb{C})$ induced from a character of $H_1$, say $\Delta(H_1)$, when
restricted to to $H_2$. 
This is given by Mackey's induction-reduction theorem in the form
\begin{eqnarray}\label{eq7}
\big(\Delta(H_1)\uparrow SL(2,\mathbb{C})\big)\downarrow H_2 =
\sum_{\oplus d} \Delta_d(S_d)\uparrow H_2,
\end{eqnarray}
where the direct sum is over all the $H_1 d H_2$ double cosets in
 $SL(2,\mathbb{C})$ of non-zero measure. This reduction is found in the following steps.

{\bf Step 1.} The double cosets $H_1 d H_2$ are found as follows. 
Naimark has shown \cite{Naimark} that the right coset representatives of $H_1$ in $SL(2,
\mathbb{C})$ can be taken to be of the form
\begin{eqnarray}\label{eq8}
Z=\left(\begin{array}{cc}1&0\\z&1\end{array}\right),\qquad z = x+iy\in \mathbb{C}.
\end{eqnarray}
The double coset representatives can be computed by
\begin{eqnarray*}
&&\left(\begin{array}{cc}1&0\\z'&1\end{array}\right)
\left(\begin{array}{cc}1&\beta\\0&1\end{array}\right)
=
\left(\begin{array}{cc}\lambda&\mu\\0&\lambda^{-1}\end{array}\right)
\left(\begin{array}{cc}1&0\\z&1\end{array}\right)
\\ &&\Rightarrow
\left(\begin{array}{cc}1-\beta z&\beta\\z'-z-\beta zz'&\beta z'+1\end{array}\right) =
\left(\begin{array}{cc}\lambda&\mu \\ 0 &\lambda^{-1}\end{array}\right).
\end{eqnarray*}
From the bottom left element of the matrix we find
\begin{eqnarray}\label{eq9}
z' = \frac z{1-\beta z}.
\end{eqnarray}
If $z'=0$, then $z=0$. This means that the origin in the complex 
plane $\mathbb{C}$ is a double coset itself. This corresponds to $d = \left(\begin{array}
{cc}1&0\\0&1\end{array}\right)$.

Next, suppose $z' = 1$. It follows from (\ref{eq9}) that
\begin{eqnarray*}
\beta = \frac1z-1,
\end{eqnarray*}
which shows that for every $\beta\in\mathbb{C}$ there exists a curve
passing through $z'=1$ and joining $z$ and $\beta$. Thus $\mathbb{C}\setminus\{0\} $ is 
another double coset, a representative of which may be taken to be any matrix
of the form $ \left(\begin{array}{cc}1&0\\z&1\end{array}\right),\ z\neq 0$.

We may choose $z=1$ and thus we have the following double coset decomposition
\begin{eqnarray*}
SL(2,\mathbb{C}) =H_1 \left(\begin{array}{cc}1&0\\0&1\end{array}\right) H_2
\cup H_1 \left(\begin{array}{cc}1&0\\1&1\end{array}\right) H_2.
\end{eqnarray*}
However, the double coset $H_1
\left(\begin{array}{cc}1&0\\0&1\end{array}\right) H_2$, 
corresponding to the origin, has measure zero. Thus there is only one double
coset of non-zero measure, {\it i.e.},
\begin{eqnarray*}
SL(2,\mathbb{C}) = H_1 \left(\begin{array}{cc}1&0\\1&1\end{array}\right) H_2,
\qquad\mathrm{almost\ \ everywhere}.
\end{eqnarray*}

{\bf Step 2.} Next, consider the subgroup
\begin{eqnarray*}
S_d = H_1\cap d^{-1} H_2 d,
\end{eqnarray*}
where $d = \left(\begin{array}{cc}1&0\\1&1\end{array}\right)$.
\begin{eqnarray*}
d^{-1} H_2 d = \left(\begin{array}{cc}1&0\\-1&1\end{array}\right)
\left(\begin{array}{cc}1&\beta\\0&1\end{array}\right)
\left(\begin{array}{cc}1&0\\1&1\end{array}\right)
= \left(\begin{array}{cc}1+\beta&\beta\\-\beta &1-\beta\end{array}\right),\\
S_d = \left(\begin{array}{cc}\lambda&\mu\\
0&\lambda^{-1}\end{array}\right)\cap\left(\begin{array}{cc}1+\beta& 
\beta\\-\beta&1-\beta\end{array}\right)= \left(\begin{array}{cc}1&0\\0&1\end{array}\right) = \{e\},
\end{eqnarray*}
which is just the identity element. Thus $S_d$ is the trivial subgroup of $SL(2,\mathbb{C})$. 

By Mackey's theorem, we have
\begin{eqnarray}\label{eq10}
\big(\Delta(H_1)\uparrow SL(2,\mathbb{C})\big)\downarrow H_2
= {\bf 1}(e)\uparrow H_2,
\end{eqnarray}
where ${\bf 1}(e)$ is the 1-dimensional representation of the identity
subgroup. Clearly, the right-hand side of (\ref{eq10}) is just the regular representation of $H_2$.

{\bf Step 3.} The quasi-invariant measure on $H_1\setminus SL(2,\mathbb{C})$.
As mentioned earlier the right coset representatives of $H_1$ in
$SL(2,\mathbb{C})$ may be taken to be of the form (\ref{eq8}). Let 
\begin{eqnarray*}
g = \left(\begin{array}{cc} g_{11}&g_{12}\\g_{21}& g_{22}\end{array}\right) \in
SL(2,\mathbb{C}),\qquad Z' = \left(\begin{array}{cc} 1&0\\z' &1\end{array}\right)
\end{eqnarray*}
be such that
\begin{eqnarray}
Zg\in H_1 Z'\quad\Rightarrow\quad Zg = hZ'
\end{eqnarray}
for some $h\in H_1$. Thus we have
\begin{eqnarray}\label{eq12}
\left(\begin{array}{cc} 1&0\\z& 1\end{array}\right)\left(\begin{array}{cc}
g_{11}&g_{12}\\g_{21}& g_{22}\end{array}\right)&=&
\left(\begin{array}{cc} \lambda&\mu\\0& \lambda^{-1}\end{array}\right)
\left(\begin{array}{cc}  1& 0\\z'& 1\end{array}\right)\nonumber
\\
\left(\begin{array}{cc} g_{11}&g_{12}\\zg_{21}+g_{22}& zg_{12}+g_{22}\end{array}\right)
&=&\left(\begin{array}{cc} \lambda+\mu z'&\mu\\
\lambda^{-1}z'& \lambda^{-1}\end{array}\right)\nonumber
\\
z'=\frac{\lambda^{-1}z'}{\lambda^{-1}} &=& \frac{zg_{21}+g_{22}}{zg_{12}+g_{22}}
\end{eqnarray}
Putting $z= x+iy,\ z' = x'+iy'$ in (\ref{eq12}) and separating 
the real and imaginary parts, one obtains
\begin{eqnarray}
dx' dy' = \vert zg_{12}+ g_{22}\vert^{-4}dx dy.
\end{eqnarray}
Hence, the corresponding $\lambda_\rho$-function \cite{Assar} is 
\begin{eqnarray}\label{eq14}
\lambda_\rho(H_1z,g) = \vert zg_{12}+ g_{22}\vert^{-4}.
\end{eqnarray}

{\bf Step 4.} The quasi-invariant measure on $H_2/S_d$. The
$\lambda_\rho$-function corresponding to the quasi-invariant measure on $H_2/S_d$ is 
\begin{eqnarray*}
\lambda_\rho(S_dZ',g) = \lambda_\rho(H_1(dZ'),g),
\end{eqnarray*}
where $Z' = \left(\begin{array}{cc} 1& z'\\0& 1\end{array}\right) \in H_2$ and $g\in H_2$.
\begin{eqnarray*}
&&dZ' = \left(\begin{array}{cc} 1& 0\\1& 1\end{array}\right) 
\left(\begin{array}{cc} 1& z'\\0& 1\end{array}\right)
= \left(\begin{array}{cc} 1& z'\\1& 1+z'\end{array}\right) = k,\\
&&\lambda_\rho(H_1(dZ'),g) = \lambda_\rho(H_1k,g),\quad g\in H_2.
\end{eqnarray*}
We may decompose $k$ as
\begin{eqnarray}\label{eq15}
&&k = h_1Z,\qquad h_1 \in H_1,\quad Z =\left(\begin{array}{cc} 1&0\\z& 1\end{array}\right):\nonumber\\
&&\left(\begin{array}{cc} 1&z'\\1& 1+z'\end{array}\right)=
\left(\begin{array}{cc} \lambda& \mu\\ 0& \lambda^{-1}\end{array}\right)
\left(\begin{array}{cc} 1&0\\z& 1\end{array}\right)=
\left(\begin{array}{cc} \lambda+\mu z& \mu\\\lambda^{-1}z& \lambda^{-1}\end{array}\right)
\nonumber\\&&\Rightarrow
z = \frac{\lambda^{-1}z}{\lambda^{-1}} = \frac1{1+z'}.
\end{eqnarray}
Taking $g=\left(\begin{array}{cc} 1&\beta\\0& 1\end{array}\right)\in H_2$, {\it i.e.}
 $g_{12} = \beta,\ g_{22} = 1$, we have, using (\ref{eq14}), that
\begin{eqnarray}
\lambda_\rho(H_1k,g) = \left\vert \frac{\beta}{1+z'}+1\right\vert^{-4},
\end{eqnarray}
and
\begin{eqnarray*}
dx' dy' = \left\vert\frac{1+z'+\beta}{1+z'}\right\vert^{-4}dx dy
\end{eqnarray*}
is the required quasi-invariant measure on $H_2/S_d$.

{\bf Step 5.} It follows that the unitary representation in (\ref{eq7}) is given by
\begin{eqnarray}
U(g)f(Z') =\sqrt{\left\vert\frac{1+z'+\beta}{1+z'}\right\vert^{-4}} f(Z'g),
\quad Z', g \in H_2,\quad g = \left(\begin{array}{cc} 1&\beta\\0& 1\end{array}\right).
\end{eqnarray}
Using
\begin{eqnarray*}
\left(\begin{array}{cc} 1& z'\\0& 1\end{array}\right)
\left(\begin{array}{cc} 1&\beta\\0& 1\end{array}\right)=
\left(\begin{array}{cc} 1&z'+\beta \\0 & 1\end{array}\right),
\end{eqnarray*}
identifying
\begin{eqnarray*}
\left(\begin{array}{cc} 1& z'\\0& 1\end{array}\right) \leftrightarrow z',\quad \mathrm{etc.}
\end{eqnarray*}
and finally changing $z'$ to $z$, we obtain
\begin{eqnarray}\label{eq18}
U(\beta)f(z) = \left\vert \frac{1+z}{1+z+\beta}\right\vert^2 f(z+\beta).
\end{eqnarray}
This is a unitary representation of $H_2$, obtained by the restriction of a
representation $\Delta(H_1) \uparrow SL(2,\mathbb(C))$ to the subgroup $H_2$.

{\bf Step 6.} Determination of the infinitesimal operators ({\it i.e.}
the generalized Lie derivatives). The generators of $H_2$ are 
\begin{eqnarray*}
b_1 = \frac d{d\beta_1}\left(\begin{array}{cc} 1&\beta_1+i\beta_2\\0& 1\end{array}\right)
\Big\vert_{\beta = 0} = \left(\begin{array}{cc} 0&1\\0& 0\end{array}\right),\\
b_2 = \frac d{d\beta_2}\left(\begin{array}{cc} 1&\beta_1+i\beta_2\\0& 1\end{array}\right)
\Big\vert_{\beta = 0} = \left(\begin{array}{cc} 0&i\\0& 0\end{array}\right).
\end{eqnarray*}
The corresponding 1-parameter subgroups are 
\begin{eqnarray*}
g_1 = \exp\{\beta_1 b_1\},\qquad g_2 =\exp\{\beta_2 b_2\}.
\end{eqnarray*}
The associated infinitesimal operators are given by the Lie derivatives
\begin{eqnarray*}
B_i = \lim_{\eta\rightarrow 0} \frac{U(\exp(\eta b_i))-U(e)}{\eta},\quad i=1,2\ .
\end{eqnarray*}
That is
\begin{eqnarray*}
B_1 f(z) = \lim_{\beta_1\rightarrow 0} \frac{U(\beta_1)f(z)-U(0)f(z)}{\beta_1},\\
B_2 f(z) = \lim_{\beta_2\rightarrow 0} \frac{U(i\beta_2)f(z)-U(0)f(z)}{\beta_2}.
\end{eqnarray*}
We compute $B_1$. Writing $f(z) = f(x,y)$, where $z=x+iy$, we have

\begin{eqnarray}\label{eq19}
&&U(\beta_1)f(z) = U(\beta_1) f(x,y) = 
\left\vert\frac{z+1}{1+z+\beta_1}\right\vert^2 f(x+\beta_1,y)\nonumber\\
&&U(0) f(z) = f(z) = f(x,y)\nonumber\\
&&B_1 f(z) = \lim_{\beta_1\rightarrow 0} \frac{
\vert z+1\vert^2 \frac{f(x+\beta_1,y)}{\vert 1+z+\beta_1\vert^2}-f(x,y)}{\beta _1}.
\end{eqnarray}
Let 
\begin{eqnarray*}
\frac{f(x+\beta_1,y)}{\vert 1+z+\beta_1\vert^2} = 
\frac{f(x+\beta_1,y)}{(1+x+\beta_1)^2 + y^2} =: F(x+\beta, y).
\end{eqnarray*}
Expanding this function in a Taylor series
\begin{eqnarray*}
&&F(x+\beta, y) = F(x,y) +\beta_1\frac\partial{\partial x} F(x+\beta_1,y)\bigg\vert_{\beta_1=0}
+\dots\\&&
= \frac{f(x,y)}{\vert z+1\vert^2}\\&& + 
\beta_1\frac{
\vert z+1+\beta_1\vert^2\frac{\partial}{\partial x}f(x+\beta_1,y)
-2(x+1+\beta_1)f(x+\beta_1, y)}{\vert z+\beta_1+1\vert^4}\bigg\vert_{\beta_1=0}+\dots
\end{eqnarray*}
Replacing this in (\ref{eq19}) we obtain
\begin{eqnarray*}
B_1 f(z) = \left\{\frac{-2(x+1)}{\vert z+1\vert^2}+\frac\partial{\partial x}\right\}f(z),
\end{eqnarray*}
and hence we have
\begin{eqnarray}
B_1 = \frac{-2(x+1)}{(x+1)^2+y^2}+\frac\partial{\partial x}.
\end{eqnarray}
A similar computation yields 
\begin{eqnarray}
B_2 = \frac{-2y}{(x+1)^2+y^2}+\frac\partial{\partial y}.
\end{eqnarray}
Applying a change of variable, $x+1 \rightarrow x$, one obtains a more symmetric form,
\begin{eqnarray}\label{eq22}
B_1 = \frac{-2x}{x^2+y^2}+\frac\partial{\partial x},\quad 
B_2 = \frac{-2y}{x^2+y^2}+\frac\partial{\partial y}.
\end{eqnarray}
Note that $[B_1,B_2] = 0$ as one expects.

{\bf Step 7}. The measure which makes the space of functions $f(z)$ in
(\ref{eq18}) into a Hilbert space $L^2(-\infty,+\infty)$ is the same measure
with respect to which the representation $U(\beta)$ is unitary,
\begin{eqnarray*}
\big(U(\beta) f_1(z), U(\beta) f_2(z)\big) = \big(f_1(z),f_2(z)\big).
\end{eqnarray*}
This measure is obtained from relation (\ref{eq15}),
\begin{eqnarray*}
&&z = \frac1{1+z'} = \frac{(1+x')-iy'}{(1+x')^2+y'^2} \Rightarrow\\
&&x= \frac{1+x'}{(1+x')^2+y'^2},\quad y=\frac{-y'}{(1+x')^2+y'^2}\\
&&dxdy = \left\vert\begin{array}{cc} \frac{\partial x}{\partial x'} &
\frac{\partial y}{\partial x'}\\ \frac{\partial x}{\partial y'}&
\frac{\partial y}{\partial y'}\end{array}\right\vert dx'dy' = 
\frac{dx'dy'}{\vert 1+z'\vert^4},
\end{eqnarray*}
and if the same change of variables, $x+1\rightarrow x,\ y\rightarrow y$ 
is applied, one obtains
\begin{eqnarray*}
dxdy = \frac1{\vert z'\vert^4} dx'dy' = \frac{dx' dy'}{(x'^2+y'^2)^2}.
\end{eqnarray*}
The $L^2(-\infty,+\infty)$ space of functions $f(z)$ has a product given by
\begin{eqnarray}
\big(f_1(z), f_2(z)\big) = \int_{-\infty}^{+\infty}\int_{-\infty}^{+\infty} 
\overline{f_1(z)}f_2(z)\frac{dx dy}{\vert z\vert^4}.
\end{eqnarray}
It is easily seen that this is the product with respect to which $U(\beta)$ is unitary.

With proper behaviour of $f_1,\ f_2$ at infinity, the operators $B_1$ and
$B_2$ given by (\ref{eq22}) are antisymmetric,
\begin{eqnarray*}
(B_if, g) = -(f, B_ig),\quad i=1,2.
\end{eqnarray*}
Therefore, the operators
\begin{eqnarray}
\Pi_1 = iB_1 = -\frac{2ix}{x^2+y^2}+i\frac\partial{\partial x},\nonumber\\
\Pi_2 = iB_2 = -\frac{2iy}{x^2+y^2}+i\frac\partial{\partial y},\\
\end{eqnarray}
are self-adjoint and their commutator vanishes. As these operators 
correspond to translations in the hyperplane introduced earlier, they are the momentum 
operators acting on the wave function of the localized massless particle.

Introducing the operator
\begin{eqnarray}
P^2:= \Pi_1^2+\Pi_2^2
\end{eqnarray}
and using the fact that $[P^2,\Pi_1]=0=[P^2,\Pi_2]$, the simultaneous
 eigenfunctions of $\Pi_1$ and $\Pi_2$ can be computed from
\begin{eqnarray}
P^2 \varphi = \lambda^2\varphi.
\end{eqnarray}
The computation of $\varphi$ is easily done by writing the expression
 for $P^2$ in polar coordinates corresponding to the cartesian coordinates $(x,y)$. W
e propose that the wave function corresponding to the localized massless 
state is a superposition of all such eigenstates.

\end{document}